\documentclass[aps,prd,amsmath,twocolumn,showpacs,letterpaper,superscriptaddress]{revtex4}
 \usepackage{amssymb}

\usepackage{graphicx}
\usepackage{psfrag}
\usepackage{color}

\newcommand{\comment}[1]{}
\usepackage{mathptmx}

\begin{document}

\title{Photons with sub-Planckian Energy Cannot Efficiently Probe Space-Time Foam}
\author{Yanbei Chen}
\email{yanbei@its.caltech.edu }
\affiliation{Theoretical Astrophysics 350-17, California Institute of Technology, Pasadena, CA 91125, USA}
\author{Linqing Wen}
\email{linqing.wen@uwa.edu.au}
\author{Yiqiu Ma}
\email{myqphy@gmail.com}
\affiliation{Australian International Gravitational Research Centre,
School of Physics, University of Western Australia, 35 Stirling Hwy,
Crawley, WA 6009, Australia}
\date{\today}

\begin{abstract}
Extra-galactic sources of photons have been used to constrain space-time quantum fluctuations in the Universe.  In
these proposals, the fundamental ``fuzziness'' of distance caused by
space-time quantum fluctuations has been directly identified with
fluctuations in optical paths. Phase-front corrugations deduced from
these optical-path fluctuations are then applied to light from
extra-galactic point sources, and used to constrain various models of
quantum gravity.  However, when a photon propagates
in three spatial dimensions, it does not follow a specific ray, but
rather samples a finite, three-dimensional region around that
ray --- thereby averaging over space-time quantum fluctuations all
through that region.  We use a simple, random-walk type model to
demonstrate that, once the appropriate wave optics is applied, the
averaging of neighboring space-time fluctuations will cause much
less distortion to the phase front. In our model, the extra
suppression factor due to diffraction is the wave length in units of the Planck length, which
is at least $10^{29}$ for astronomical observations.
 \end{abstract}

 \pacs{95.75.Kz, 03.67.Lx, 04.60.-m, 98.62.Gq}
\maketitle

\section{Introduction}
There have been several proposals to use extra-galactic point sources to
constrain the quantum fluctuations in space-
time~\cite{NCvD,LH,RTG,CNvD}. It was argued that, space-time
fluctuations cause random phase shifts of photons, and that these
shifts accumulate throughout the very long light propagation
path from the point source to the earth, causing wavefront
distortion from a perfect spherical shape upon arrival at the earth.
The manner in which these fluctuations accumulate depends on the specific
model of quantum gravity phenomenology; and in particular, it has been claimed
that the {\it random walk} model could be ruled out by existing
imaging data from the Hubble Space Telescope. In this paper, we
point out  a serious omission in the theory so far
employed by all such proposals, and argue that the random walk model,
once given a closer look, cannot be ruled out at all by current or
any forseeable observations of extra galactic sources of photons.

In Refs.~\cite{NCvD,LH,RTG,CNvD}, based on the argument about the
quantum space-time on the phenomenological level~\cite{Amelino,NCvD},
the authors assumed photons
originating from a point source to undergo a random phase shift
$\Delta \phi$ due to space-time fluctuations:
\begin{equation}
\label{eqphi}
\Delta \phi  \sim 2\pi (l_{\rm P}/\lambda)^{\alpha} (L / \lambda)^{1-\alpha},
\end{equation}
 where $l_{\rm P}$ is the Planck length, $\lambda$ is the wavelength of the light, $L$ is the light propagation length, and $\alpha$ is a  parameter that depends on specific models of quantum space-time phenomenology. In particular, $\alpha=1/2$ corresponds to the so-called {\it random-walk model}, which can be understood as having the speed of light fluctuating dramatically at the scale of $l_{\rm P}$.   The relation~\eqref{eqphi} stems from phenomenological description of the quantumness of space-time by assuming that the uncertainty in distance measurement $\delta L$ due to space-time fluctuation over a distance $L$ is given by
 \begin{equation}
\label{eql}
 \delta L \ge L^{1-\alpha} l_{\rm P}^{\alpha}\,,
 \end{equation}
 which was first derived by Ng and van Dam  and then discussed by others~\cite{CNvD,SciAm,GLM,MM}.
 Note that Eq.~\eqref{eqphi} is related to Eq.~\eqref{eql} by  $\Delta \phi  \sim 2\pi (\delta L /\lambda)$.

If we consider a photon propagating along one spatial dimension.  Suppose we divide its propagation distance $L$ into pieces of $l_{\rm P}$; within each piece, the phase-shift fluctuation it gains is substantial: $\delta \phi
\sim  2\pi l_{\rm P}/\lambda $ --- while fluctuations in different
intervals are independent of each other. In this way, the total
phase shift  of the photon does a ``random walk'' while the light
propagates.  At the end of propagation, we have a photon-phase
fluctuation of
\begin{equation}
\label{phinaive}
(\Delta\phi)_{\rm 1D} \sim \sqrt{N_L} \delta\phi \sim \sqrt{l_{\rm P} L /\lambda^2}\,,
\end{equation}
which is exactly Eq.~\eqref{eqphi} with $\alpha=1/2$.

However, controversies about the magnitude of this wave-front distortion effect were raised in several papers. Coule ~\cite{Coule2003} first qualitatively argued that this decoherence
effect would be very small because of the van Cittert-Zernike theorem~\cite{vCZ}. Later, by modeling the
discrete space-time in a Lorentz-invariance way, Dowker et.al ~\cite{Dowker2010} calculated a simple model describing the electromagnetic potential generated by
oscillating charges at the source rather than studying the independent dynamics of the light wave field.
They found that the signal due to space-time foam would be undetectably tiny.

In our paper, we focus on the independent dynamics of light wave fields in space-time with Planck-scale fluctuations
by quantitatively studying a model. Actually the light field, as a wave, does not propagate in the way suggested in ~\cite{NCvD}, but in the following Huygens-Fresnel way: when a photon travels through a
space-time region, it does not follow only one particular ray, whose
length might be subject to the fundamental ``fuzziness'' prescribed by
Eq.~\eqref{eql},  but instead, the {\it wave nature of light}, or the {\it quantum mechanical nature of the photon}, dictates that the photon  would {\it simultaneously} sample
{\it an ensemble} of many different neighboring rays, each of which
has a {\it potentially different} realization of the fundamental
length fluctuation; the actual path-length fluctuation must then be
given by an {\it averaging} among these different length
fluctuations.   This allows $\Delta\phi $ to go below $2\pi \delta
L/\lambda$.  Moreover, because the linear size of the sampling
region can be much bigger than $l_{\rm P}$ (the correlation length
of fundamental quantum fluctuations), this averaging can dramatically
suppress the actual $\Delta \phi$ from  $2\pi \delta L/\lambda$, or
Eq.~\eqref{eqphi}.

In other (simpler) words: (i) propagation of photons is described by the photons' wavefunction, which describes electromagnetic (EM) waves; (ii) diffraction of EM waves makes them insensitive to  fluctuations and disturbances at scales much less than the wavelength; and (iii) as an EM wave propagates through a large distance, the coherence level of its phase front increases unless further disturbance keeps coming~\cite{vCZ} (This point was also independently discussed qualitatively by D. H. Coule ~\cite{Coule2003}).  If we divide 3-D space into cubes of side length $\sim l_{\rm P}$, with disturbance to light propagation independent within each cube, then using Fourier optics, it is easy to estimate that  the effect we have described above will be suppressed by $\sqrt{l_{\rm P}/\lambda}$ in each transverse direction (because only perturbations with spatial frequencies below $\sim 1/\lambda$ along each transverse direction get registered by the propagating light), which leads to
\begin{equation}
\label{eqphi3d}
\Delta\phi_{\rm 3D} = \Delta\phi_{\rm 1D} (l_{\rm P}/\lambda).
\end{equation}
In the following, before discussing the consequences of this rather dramatic suppression, we shall justify Eq.~\eqref{eqphi3d} at a very pedagogical level.

\smallskip

\section{Model of space-time fluctuation}
Let us construct a toy model reflecting the effect of quantum space-time fluctuation induced length uncertainty on the propagation
of light. In this model, the Minkowski space-time with length fluctuation at each point is viewed as a ``medium" with a random (yet static)
distribution of refractive index $n(x,y,z) \equiv
1+\epsilon(x,y,z)$.  We assume  the following
translational invariant spatial auto-correlation function for $\epsilon$:
\begin{eqnarray}
\label{eqscart}
&&\langle \epsilon(x',y',z') \epsilon(x'',y'',z'')\rangle  \nonumber\\
&=& a^2 \Pi(x'-x'') \Pi(y'-y'')  \Pi(z'-z''),
\end{eqnarray}
where  ``$\langle \ldots \rangle$'' stands for ensemble average,  $a$ is of order unity, and
\begin{equation}
\Pi(x)=\left\{
\begin{array}{ll}
1 & \displaystyle |x| \stackrel{<}{_\sim} l_{\rm P} \\
0 & |x| \gg l_{\rm P}.
\end{array}
\right.
\end{equation}

Our purpose is to study the propagation of light as a scalar wave traveling
in this medium. In this specific toy model, the coordinate speed of light propagation
fluctuates, in on a small region with the size
comparable to the Planck scale --- this simulates light propagation
in a space-time with Planck-scale quantum fluctuations. In addition, light-speed
fluctuations in regions separated by more than the Planck length are
independent of each other following the random
walk model. As we shall see in the calculation below, the particular shape of the correlation function Eq.~\eqref{eqscart} does not matter --- the effect will remain the same as long as: (i) the total variance in $n$ is of order unity, and (ii) coherence in $\epsilon$ exists only between points separated by less than  than $l_{\rm p}$. (3) Our toy model entails a quantum gravity induced breaking of Lorentz invariance which has been strongly restricted by astrophysical data~\cite{Amelino}. However, our toy model should be sufficient to demonstrate the omitted effect in \cite{NCvD,LH,RTG,CNvD} due to the wave nature of light and to capture the key character of the previously proposed idea.

\smallskip

\section{The Wave Equation}
Returning to the wave picture, we first write down the wave equation:
\begin{equation}
-[1+2\epsilon(x,y,z)]\frac{\partial^2\Phi}{\partial t^2} + \nabla^2 \Phi =0\,.
\end{equation}
Since our refractive-index perturbation is static, we can expand the
total wave into two monochromatic pieces, the ideal wave
$\Phi_0(x,y,z) e^{-i\omega_0 t}$ and the scattered wave
$\psi(x,y,z) e^{-i\omega_0 t}$:
\begin{equation}
\Phi = \left[\Phi_0(x,y,z) + \psi(x,y,z) \right]e^{-i\omega_0 t}.
\end{equation}
The ideal wave is the unperturbed part of the light field satisfying:
$-\partial^2\Phi_0/\partial t^2+\nabla^2\Phi_0=0$.
At leading order in $\epsilon$, we
have
\begin{equation}
\label{perturbedwe}
\left(\nabla^2 +\omega_0^2 \right) \psi(x,y,z) = - 2\omega_0^2 \epsilon(x,y,z)  \Phi_0(x,y,z)\,,
\end{equation}
which means $\psi$ is a perturbative field sourced by a beat between the ideal wave and space-time perturbations.  We know from the observed phenomenology that such a perturbation must apply to our situation: namely, fluctuation caused by the space-time foam is indeed very weak compared with an ideal wave propagating across the Universe, and we have $|\psi /\Phi_0| \ll 1$.

For a point source, we assume
\begin{equation}
\Phi_0 (x,y,z) = \frac{e^{i\omega_0 r}}{4\pi r} \,,\quad r \equiv \sqrt{x^2+y^2+z^2}\,.
\end{equation}
At the distance $L$, the scattered wave $\psi$ must be compared with
the ideal wave to characterize the modulation caused to the idea spherical wave by space-time perturbations. Let us define
\begin{equation}
\alpha + i\phi \equiv \frac{\psi}{\Phi_0} =  4\pi L \psi e^{-i\omega_o L}\,,\quad \alpha,\phi \in \mathbb{R}\,,
\end{equation}
so $\alpha$ describes the amplitude modulation, and $\phi$ describes phase modulation in radians. We also define the {\it total
modulation},
\begin{equation}
\label{xidef}
\xi  \equiv \sqrt{\langle \alpha^2 + \phi^2\rangle} = 4\pi  L  \sqrt{\langle \psi \psi^* \rangle} \,,
\end{equation}
whose standard deviation is greater than those of both the amplitude and the phase modulations.

\smallskip

\section{Summing Over Paths}
To arrive at the answer quickly, we use
the Huygens-Fresnel-Kirchhoff scalar diffraction theory, which is
equivalent to applying the outgoing Green Function, and
obtain~\cite{jackson}:
\begin{eqnarray}
\label{green}
\psi(\mathbf{x})\!\! &=&\!\! \int_{|\mathbf{x}'|<L}  -2\omega_0^2 \epsilon(\mathbf{x}')\Phi_0(\mathbf{x}') \frac{e^{i\omega_0 |\mathbf{x} - \mathbf{x}'|}}{4\pi |\mathbf{x} - \mathbf{x}'|}d\mathbf{x}' \nonumber \\
\!\!&=&\!\!\int_{|\mathbf{x}'|<L} \frac{e^{i\omega_0 |\mathbf{x'}|}}{4\pi|\mathbf{x}'|} [-2\omega_0^2\epsilon(\mathbf{x}')]\frac{e^{i\omega_0 |\mathbf{x} - \mathbf{x}'|}}{4\pi |\mathbf{x} - \mathbf{x}'|}d\mathbf{x}'.\;
\end{eqnarray}
Note that we have considered only contributions from fluctuations at
distances smaller than $L$ to the point source. The
integral~\eqref{green} can be interpreted as a path integral ---
over all paths that consist of two straight sections (each
associated with a propagator), and a deflection in the middle due to
interaction with refractive-index fluctuations (associated with a
coupling coefficient).  Paths with more than one deflection do not
have to be taken into account in our perturbative treatment at leading order.

If we discretize the integration domain, a sphere with volume $\sim L^3$, into cells with linear size $\sim l_{\rm P}$ and volume
$v_{\rm P} \sim l_{\rm P}^3$, we will get a total of  $N_{\rm tot}
\sim L^3/l_{\rm P}^3$ individual cells, each of which has a statistically
independent fluctuation in $\epsilon$ with variance $a^2$
[cf.~Eq.~\eqref{eqscart}]. Then the fluctuation variance of $\psi$ given in Eq.~\eqref{green}
can be estimated in the following way:
\begin{equation}\label{variance}
\langle |\psi |^2\rangle\sim \omega_0^4 \int d\mathbf{x}d\mathbf{x'}
\frac{e^{i\omega_0( |\mathbf{x'}|+ |\mathbf{x} - \mathbf{x}'|-|\mathbf{x''}|-|\mathbf{x} - \mathbf{x}''|)}}{|\mathbf{x} - \mathbf{x}''||\mathbf{x} - \mathbf{x}'||\mathbf{x'}||\mathbf{x''}|}
\langle \epsilon(\mathbf{x'})\epsilon(\mathbf{x''})\rangle,
\end{equation}
where the integration is over the region $|\mathbf{x}|<L,|\mathbf{x'}|<L$.
Since the correlation function of $\epsilon(\mathbf{x})$ only contributes to the integral when $|\mathbf{x'}-\mathbf{x''}|\leq|l_P|$, therefore the exponential in the numerator is approximately equal to 1. If then we change the integration argument using $\Delta \mathbf{x}=\mathbf{x'}-\mathbf{x''}$ and $2\mathbf{x_0}=\mathbf{x'}+\mathbf{x''}$ while substituting Eq. \eqref{eqscart}, the variance becomes:
\begin{equation}
\langle |\psi |^2\rangle\sim \omega_0^4\int_{|\Delta \mathbf{x}|<l_P}d^3\Delta\mathbf{x}\int_{|\mathbf{x}_0|<L}
d^3\mathbf{x}_0\frac{1}{|\mathbf{x}_0|^2|\mathbf{x}-\mathbf{x}_0|^2}.
\end{equation}
The first integral on $\Delta\mathbf{x}$ is just $l_P^3$ while the second one on $\mathbf{x}_0$ can be approximated to be $L^3/L^4$. Using the definition of $v_P$ and $N_{\text{total}}$, finally we have:
\begin{equation}
\sqrt{\langle |\psi |^2}\rangle \sim \frac{ \omega_0^2 a}{L^2} v_{\rm P} \sqrt{N_{\rm tot}} \sim a\omega_0^2 \sqrt{l_{\rm P}^3/L}\,.
\end{equation}

According to Eq.~\eqref{xidef} and comparing with Eq.~\eqref{phinaive}, we have
\begin{equation}
\label{phiorder}
\Delta\phi \stackrel{<}{_\sim}\xi \sim a\omega_0^2 \sqrt{l_{\rm P}^3 L} \sim (\Delta\phi)_{\rm 1D} (l_{\rm P}/\lambda)\,.
\end{equation}
There is an extra suppression factor of  $l_{\rm P}/\lambda$, which
arises from the fact that in Eq.~\eqref{green}, the intermediate
point $\mathbf{x}'$ of the optical path has the freedom to depart away
from the axis connecting the source point and the field point, and
sample through $N_{\rm tot} \sim L^3/l_{\rm P}^3$ independent
fluctuations, instead of only $N_L \sim L /l_{\rm P}$ in the
one-dimensional treatment.  A more precise calculation gives only an additional numerical factor of the order of unity:
\begin{equation}
\label{xikirch}
\Delta\phi \stackrel{<}{_\sim}\xi = \sqrt{\frac{\pi}{8}} a\omega_0^2 \sqrt{ l_{\rm P}^3 L}  \,.
\end{equation}
We have therefore confirmed Eq.~\eqref{eqphi3d}.

\smallskip

\section{Spatial-Scale Cut-off}
To separately study fluctuations at different
spatial scales, we solve the same problem by decomposing the
secondary wave into modes:
\begin{equation}
\psi(r,\theta,\phi) = \sum_{\ell m} [\psi_{\ell m}(r) Y_{\ell  m}(\theta,\varphi)].
\end{equation}
Here $Y_{\ell m}(\theta,\varphi)$ are spherical harmonics. They
describe angular variations at scales of $2\pi/\ell $; at a radius
$r$, that corresponds to transverse length scales of $2\pi r/\ell $,
or transverse spatial frequencies of $\ell /(2\pi r)$.  The modal
decomposition of Eq.~\eqref{perturbedwe} is
\begin{eqnarray}
\label{sphpsi}
&&\left[\frac{1}{r}\frac{\partial}{\partial r}\left(r\frac{\partial}{\partial r}\right)  + \omega_0^2 -\frac{\ell  (\ell  +1)}{r^2}\right]\psi_{\ell  m}(r)   \nonumber \\
&=&-\frac{ \omega_0^2 e^{i \omega_0 r}\epsilon_{\ell  m}(r)}{2\pi r}\,, \quad
\end{eqnarray}
with\begin{equation}
\epsilon_{\ell  m}(r) \equiv \int_0^{2\pi} d\varphi \int_0^\pi{\sin\theta d\theta } \left[\epsilon(r,\theta,\varphi) Y^*_{\ell  m}(\theta,\varphi)\right]\,,
\end{equation}
which satisfies
\begin{equation}
\label{epslm}
\langle \epsilon_{\ell  m}(r)\epsilon^*_{\ell 'm'}(r')\rangle =  a^2 \delta_{\ell  \ell'}\delta_{mm'} \delta(r-r') l_{\rm p}^3 /r^2\,.
\end{equation}
Here we simply assumed
\begin{equation}
\label{epssph}
\langle \epsilon(\mathbf{x})\epsilon(\mathbf{x}')\rangle  = a^2 l_{\rm P}^3 \delta^{(3)}(\mathbf{x}-\mathbf{x}')\,.
\end{equation}
The spatial spectrum corresponding to this correlation function is identical to that in Eq.~\eqref{eqscart} at low spatial frequencies, but continues to exist in orders  higher than $1/l_{\rm P}$.  In principle, those modes may also add incoherently to our output fluctuations, but as we shall see, their contributions will be negligible.

\begin{figure}[t]
\psfrag{xx}{$\xi_\ell^2$}
\psfrag{ll}{$\ell  /(\omega_0 L)$}
\includegraphics[width=0.45\textwidth]{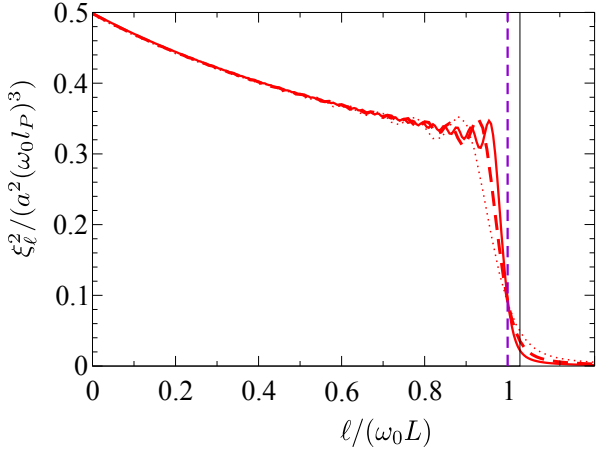}
\caption{Plots of $\xi_\ell^2$ as functions of $\ell  /(\omega_0 L)$, for cases with $\omega_0 L=100$ (dotted curve), 200 (dashed curve), and 300 (solid curve).
   \label{fig1} }
\end{figure}

Solving Eq.~\eqref{sphpsi}, assuming regularity at $r=0$ and the outgoing wave condition at $r =L$, we obtain:
\begin{equation}
\label{psimode}
\psi_{\ell m}(L) = - \frac{ \omega_0^3 L h_{\ell }^{(1)}(\omega_0 L)}{2\pi} \int_0^L dr [r j_{\ell }(\omega_0 r) e^{i\omega_0 r} \epsilon_{\ell m}(r)],
\end{equation}
where $j_\ell $ and $h^{(1)}_\ell $ are spherical Bessel
and first-kind spherical Hankel functions \footnote{In order to obtain
Eq.~\eqref{psimode}, we break the right-hand side of
Eq.~\eqref{sphpsi} into an integral over $\delta$-functions: $\int
dr' F(r)\delta(r-r')$. For each $\delta(r-r')$, its contribution
must be of the form $A j_\ell (\omega_0 r)$ for $r<r'$ (due to
regularity at origin) and $B h_\ell ^{(1)}(\omega_0 r)$ for $r>r'$
(due to outgoing condition at infinity). Then $A$ and $B$ can be
solved by applying continuity in $\psi$, and junction condition in
$\partial\psi/\partial r$.}. From Eqs.~\eqref{epslm} and~\eqref{psimode}, we obtain
\begin{eqnarray}
\label{psisphapp}
16\pi^2 L^2 \langle \psi_{\ell m} \psi_{\ell'm'}^*\rangle
= \xi_{\ell m}^2 \delta_{\ell \ell'} \delta_{mm'}\,,
\end{eqnarray}
with
\begin{eqnarray}
\label{xil}
\xi_{\ell m}^2\equiv  4 a^2 \left|\omega_0 L h_{\ell }^{(1)}(L)\right|^2  (\omega_0 l_{\rm P})^3 \int_0^L j^2_\ell (\omega_0r) \omega_0 dr \,.
\end{eqnarray}
Note that $\xi_{\ell  m}$ is independent of $m$, which is a
consequence of the rotation invariance of the refractive-index
fluctuations. The total fluctuation at $r=L$  will then be
\begin{eqnarray}
\xi^2
&=& \sum_{\ell =0}^{+\infty} \frac{2\ell +1}{4\pi}\xi_{\ell 0}^2 \equiv \sum_{\ell =0}^{+\infty} \xi_\ell ^2\,.
\end{eqnarray}
Physically, $\xi_{\ell}^2$ describes fluctuations at the angular
scale $\sim 2\pi/{\ell}$, or transverse spatial scales  $2\pi
L/\ell$, or transverse spatial frequency of $\ell /(2\pi L)$.
Inserting Eq.~\eqref{xil}, we have
\begin{eqnarray}
\label{ximodal}
{\xi_\ell^2}&=&{a^2(\omega_0 l_{\rm P})^3}  \nonumber \\
&\times&   \frac{2 \ell +1}{\pi}\left| \omega_0 L h_\ell^{(1)}(\omega_0 L)\right|^2\int_0^{\omega_0 L} j_\ell^2(R)dR.
\end{eqnarray}
We expect $\ell  \sim \omega_0 L$, or $2\pi L /\ell  \sim \lambda$
to be the turning point, because at this point the transverse
spatial scale is comparable to the wavelength $\lambda$.

Mathematically, for $\ell   < \omega_0 L $, the spherical Bessel and
Hankel functions are wavelike at $r \sim L$, indicating {\it
propagating waves}; for $\ell  > \omega_0 L$, the spherical Bessel
and Hankel functions are not wavelike at $r \sim L$, indicating {\it
evanescent waves}.   In the limiting regimes of $\ell \ll \omega_0
L$  and $\ell \gg \omega_0 L$, $\xi_\ell$   can be evaluated
analytically, using asymptotic expansions of spherical Bessel
functions:
\begin{equation}
\label{asymp}
\frac{\xi_\ell^2}{a^2 (\omega_0 l_{\rm P})^3} = \left\{
\begin{array}{ll}
1/2 \,, & \ell\ll \omega_0 L\,, \\
\omega_0 L /[(2\ell+1)^2\pi] \,, & \ell\gg \omega_0 L\,.
\end{array}\right.
\end{equation}

Note that not only does $\xi_{\ell}^2$ approach $0$ at orders $\ell
\gg \omega_0 L$, the summation of all these higher modes also gives
a negligible contribution.  This qualitatively confirms a cut-off at
the transverse scale of $\lambda$:  fluctuations at much finer
scales do not generate a secondary wave. This justifies our original
use of Eq.~\eqref{epssph}, and also ensures that the detailed shape of the correlation function~\eqref{eqscart} does not matter. In Fig.~\ref{fig1}, we study the transition zone of  $\ell \sim
\omega_0 L$ numerically, for moderately large values of $\omega_0 L
=100$, 200, and 300, by plotting $\xi_{\ell}^2$ as a function of $\ell
/ (\omega_0 L)$. As $\omega_0 L \rightarrow +\infty$, $\xi^2_{\ell}$
asymptotes to a smooth, non-zero function for $\ell/(\omega_0 L)<1$,
and to $0$ for $\ell/(\omega_0 L)>1$.  This means, in the realistic
situation of $\omega_0 L \ll 1$,  $\ell/(\omega_0 L) =1$ is a sharp
turning point between propagating and evanescent waves.

It might seem difficult to evaluate the summation \eqref{ximodal}
analytically. But since we are solving exactly the same problem as
the previous section, it should be clear that
[cf.~Eq.~\eqref{xikirch}]
\begin{equation}
\xi = \sqrt{\sum_{\ell=0}^{+\infty} \xi_{\ell}^2}  = \sqrt{\frac{\pi}{8}} a\omega_0^2 \sqrt{ l_{\rm P}^3 L}  \sim (\Delta \phi)_{\rm 1D}(l_{\rm P}/\lambda)\,,
\end{equation}
as we have verified numerically in the cases $\omega_0 L =100$, 200, and 300.


\section{Discussions}
So far in this paper, we have calculated
fluctuations on the phase front of an extra-galactic point source,
caused by Planck-scale fluctuations in refractive index --- a toy
model for space-time foam.  If diffraction of light were ignored, or
if we assumed a space-time with one time dimension plus one spatial
dimension,  our toy model  would give comparable results to previous
esimates on the random-walk model claimed in~\cite{NCvD,LH,RTG,CNvD}. However,
the diffraction of light requires us to average space-time
fluctuations over all different possible optical paths that extend
to all three spatial dimensions. This averaging filters out all
fluctuations with transverse scales finer than the wavelength. In
our model, this causes an extra suppression factor of $l_{\rm
P}/\lambda$, with   [cf.~Eq.~\eqref{eqphi}]
\begin{equation}
\label{xifinal}
\Delta\phi \stackrel{<}{_\sim}    \sqrt{l_{\rm P} L /\lambda^2}\,(l_{\rm P}/\lambda ).
\end{equation}
With respect to previous literature, this suppression is at least by 29 orders of magnitude for
astronomical observations, if the wavelength of
$\lambda=10^{-6}\,$m were to be used. Numerically, we have
\begin{equation}
\Delta\phi \sim 10^{-26} \sqrt{{L}/{\,{\rm Gpc}}}\left({5\times10^{-7}\rm{\rm m}}/{\lambda}\right)^2,
\end{equation}
which makes the quantum foam effect on the light propagation
extremely small.   Suppose we consider another type of experiment, namely the use of $\gamma$-ray arrival time. It is straightforward to convert
\begin{equation}
\Delta t =\Delta \phi \frac{\lambda}{c} = t_P \sqrt{l_{\rm P} L /\lambda^2}  = 10^{-33}\,\mathrm{s}  \sqrt{{L}/{\,{\rm Gpc}}} ({E_{\gamma}}/{\mathrm{GeV}}).
\end{equation}
For both methods, the effect of random-walk-type space-time foam is  too small to be seen.

For certain extra-dimension models~\cite{MK}, e.g, the one raised by Arkani-Hamed et. al ~\cite{ADD}, in which the fundamental scale
of nature is the electro-weak scale $l_{EM}\sim10^{-18}$m, $l_{\rm P}$ is amplified by a large factor, therefore apparently increasing the detectability of quantum foams.   Using $l_{\rm P} \rightarrow l_{EM}\sim10^{-18}\,$m, we update the above estimates to
\begin{equation}
\Delta \phi_{\rm ed}\sim 0.1 \sqrt{{L}/{\,{\rm Gpc}}}[(5\times10^{-7}\rm{\rm m})/{\lambda}]^2,
\end{equation}
and
\begin{equation}
\Delta t_{\rm ed} \sim  (10^{-8}\,{\rm s}) \, \sqrt{{L}/{\,{\rm Gpc}}} ({E_{\gamma}}/{\mathrm{GeV}}).
\end{equation}

At first sight, these seem more promising to detect.  However, we must be more careful in connecting $\Delta \phi_{\rm ed}$ and $\Delta t_{\rm ed}$, which are the {\it total variance} after summing over all spatial frequencies,  to observables in actual experiments.  In both cases, from our study of wave propagation, we have
\begin{equation}
\langle \Delta \phi
(\vec x) \Delta \phi(\vec x') \rangle \approx
\left\{
\begin{array}{cc}
\Delta\phi^2_{\rm ed}  &  |\vec x -\vec x'| \stackrel{<}{_\sim} \lambda, \\
\\
0  &|\vec x -\vec x'|> \lambda,
\end{array}\right.
\end{equation}
and a similar relation for the two-point correlation function of $\Delta t(\vec x)$\,.

For a telescope image, the ideal wave gives an Airy pattern on the focal plane, while the scattered wave would create a diffuse background on the focal plane, which has a total energy of $\Delta\phi_{\rm ed}^2$, and therefore a flux $\mathcal{F}^{Pl}_{\text{background}}$ that is
\begin{equation}
\label{ratio}
\mathcal{F}^{Pl}_{\rm background}/\mathcal{F}_{\rm image} \sim\Delta\phi_{\rm ed}^2\left(\frac{\lambda^2}{\mathcal{A}}\right)\,,
\end{equation}
compared with the typical flux of the ideal image.  It could be almost impossible to detect such a background due to confusion with other types of background. For example, we can estimate the magnitude of ~\eqref{ratio} for a typical quasar source PG2112+059, which is $\sim2.5$ Gpc away from us. According to the observational data collected by using the \emph{Hubble Space Telescope} (Wide Field Camera 3), at wavelength $\lambda\sim1.25\times10^{-6}$ m, the ratio between photon flux of the sky background and photon flux of the image is $\sim 0.7\times 10^{-3}$~\cite{NJ}. However, $\mathcal{F}^{Pl}_{\text{background}}/\mathcal{F}_{\text{image}}$ is $\sim 0.8\times 10^{-15}$, which is $10^{12}$ smaller.

For detecting arrival time of $\gamma$-ray photons,  we have to be aware that the true $\Delta t$ also depends on the averaging area of our detector --- which corresponds to the pixel size, which we also denote by $\mathcal{A}$.  As a consequence, we have
\begin{equation}
\Delta t_{\rm obs} = \Delta t_{\rm ed} \sqrt{\frac{\lambda^2}{\mathcal{A}}}\,,
\end{equation}
which is likely to gain an additional factor from $\Delta t_{\rm ed}$.  The actual time resolution of a gamma ray detector, e.g. \emph{Fermi Telescope}, is $\sim 10$ $\mu$s~\cite{Fermi}. From (36), the observed correction to the arrival time of gamma ray photons with $E_\gamma=1$ GeV and source distance $\sim 1$ Gpc is $\sim 10^{-23}$ s, which is also too small to resolve.

What has been left out in this paper is possible fluctuation in time.  If the refractive index fluctuation has a white noise spectrum up to the Planck frequency, then by observing a portion of the optical spectrum $\Delta \omega$, the effect will be suppressed further by a factor of
\begin{equation}
\sqrt{ t_{\rm P}\,\Delta\omega},
\end{equation}
which is likely to be very large as well.


\section{Acknowledgements.}
The authors would like to thank  D.G.\ Blair, Z.\ Cai, X.\ Fan, H.\ Nikolai, and C.\ Zhao for helpful discussions. We thank Ning Jiang for this help in calculating the background of quasar PG2112+059. Y. C.\ and L. W.\ were supported by the Alexander von Humboldt Foundation's Sofja Kovalevskaja
Programme (funded by the German Federal Ministry of Education and Research). We also thank Dr. Ron Burman for reading the manuscript. Y.C.\ is currently supported by NSF Grant PHY-1068881 and CAREER Grant PHY-0956189.  L.W.\ acknowledge funding supports from the Australian Research Council.  Y.M.\ is currently supported by the Australian Department of Education, Science and Training.

\end{document}